\documentclass[reprint,aps,pre]{revtex4-1}
\usepackage{graphicx}
\usepackage{mathrsfs}
\usepackage{amsmath,amssymb}
\usepackage{textcomp}
\usepackage{stmaryrd}

\newcommand{\e}{\mathrm{e}}
\newcommand{\Lambdai}{\mathit{\Lambda}}
\newcommand{\Sigmai}{\mathit{\Sigma}}
\newcommand{\Deltai}{\mathit{\Delta}}

\DeclareMathOperator{\cosec}{cosec}
\newcommand{\di}{\;\mathrm{d}}
\renewcommand{\d}[2]{\frac{\mathrm{d}#1}{\mathrm{d}#2}}
\begin{document}

\title{Elasticity and Glocality: Initiation of Embryonic Inversion in \emph{Volvox}}%

\author{Pierre A. Haas}
\author{Raymond E. Goldstein}%
\affiliation{Department of Applied Mathematics and Theoretical Physics, Centre for Mathematical Sciences, 
University of Cambridge, Wilberforce Road, Cambridge CB3 0WA, United Kingdom}
\date{\today}%
\begin{abstract}
Elastic objects across a wide range of scales deform under local changes of their intrinsic properties, yet the shapes are 
\emph{glocal}, set by a complicated balance between local properties and global geometric constraints. Here, we explore this 
interplay during the inversion process of the green alga \emph{Volvox}, whose embryos must turn themselves inside out to 
complete their development. This process has recently been shown [S. H\"ohn \emph{et al.}, \emph{Phys. Rev. Lett.} 
\textbf{114}, 178101, (2015)] to 
be well described by the deformations of an elastic shell under local variations of its intrinsic curvatures and stretches, 
although the detailed mechanics of the process have remained unclear. Through a combination of asymptotic analysis and 
numerical studies of the bifurcation behavior, we illustrate how appropriate local deformations can overcome global 
constraints to initiate inversion.
\end{abstract}
\maketitle

\section{Introduction}
The shape of many a deformable object arises through the competition of multiple constraints on the object: this competition 
may be between different global constraints, such as in Helfrich's analysis \cite{helfrich} of the shape of a red blood cell 
(where intrinsic curvature effects coexist with constrained membrane area and enclosed volume). It may also be the 
competition between local and global constraints. Such deformations, which we shall term \emph{glocal}, arise for 
example in origami patterns \cite{silverberg} (where local folds must be compatible with the global geometry). They 
are of considerable interest in the design of programmable materials \cite{bende} at macro- and microscales, where 
one asks: can a sequence of local deformations 
overcome global constraints and direct the global deformations of an object?

\begin{figure}[b]
\includegraphics{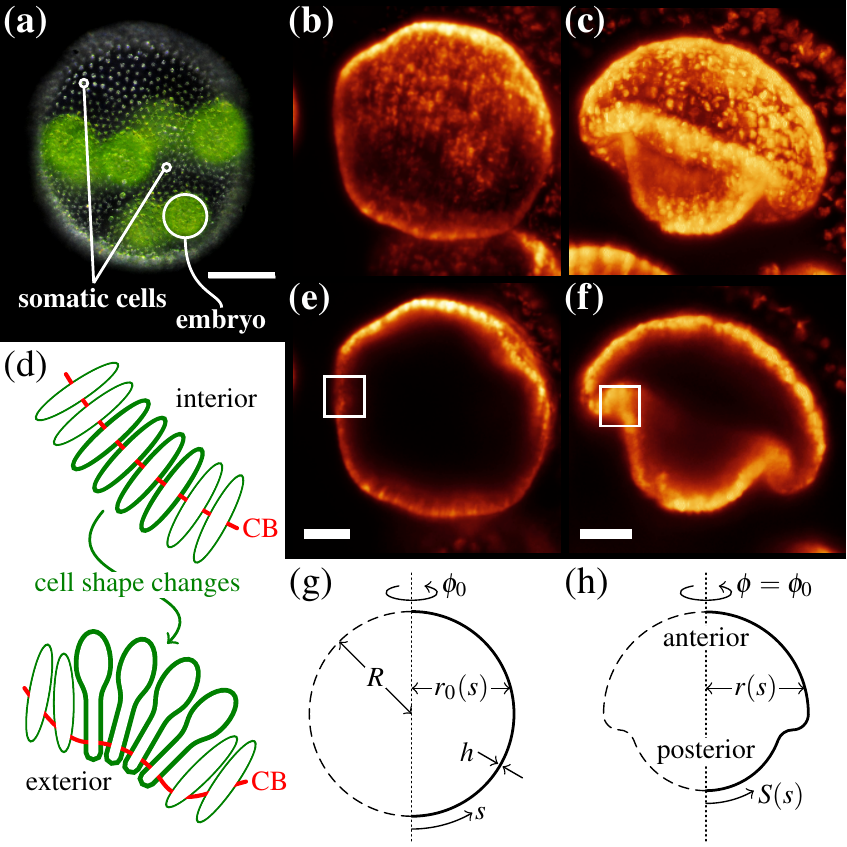}
\caption{(color online). \emph{Volvox} invagination and elastic model. (a) Adult \emph{Volvox}, with somatic cells and one
embryo labelled. (b)~\emph{Volvox} embryo at the start of inversion. (c)~Mushroom-shaped invaginated 
\emph{Volvox} embryo. (d)~Cell shape changes to wedge shapes and motion of cytoplasmic bridges (CB) bend the cell 
sheet. Red line indicates position of cytoplasmic bridges. (e,f) Cross-sections of the stages shown in panels (b,c). Cell 
shape changes as in (d) occur in the marked regions. (g)~Geometry of undeformed spherical shell of radius $R$ and 
thickness $h$. (h)~Geometry of deformed shell. Scale bars:~(a) $50\,\mbox{\textmu m}$, (e,f) $20\,\mbox{\textmu m}$. 
False color images obtained from 
light-sheet microscopy provided by Stephanie H\"ohn 
and Aurelia R. Honerkamp-Smith.}
\label{fig1}
\end{figure}

This is a problem that, at the close of their development, the embryos of the green alga \emph{Volvox} \cite{kirkbook} 
are faced with in the ponds of this world.  \emph{Volvox} (Fig.~\ref{fig1}a) is a multicellular green alga belonging to a 
lineage (the Volvocales) which has
been recognized since the time of Weismann \cite{Weismann} as  a model organism for the evolution of multicellularity, and
which more recently has emerged as the same for biological fluid dynamics \cite{ARFM}.  The
Volvocales span from
unicellular \emph{Chlamydomonas}, through organisms such \emph{Gonium}, consisting of $8$ or $16$ \emph{Chlamydomonas}-like
cells in a quasi-planar arrangement, to spheroidal species (\emph{Pandorina} and \emph{Pleodorina}) with scores or hundreds of 
cells at the surface of a transparent extracellular 
matrix (ECM).  The largest members of the Volvocales are the species of \emph{Volvox}, which display germ-soma 
differentiation, having sterile somatic cells at the surface of the ECM and a small number of germ cells in the
interior which develop to become the daughter colonies.

Following a period of substantial growth, the germ cells of \emph{Volvox} undergo repeated rounds of cell division, at the end of which
each embryo (Fig.~\ref{fig1}b,e) consists of a few thousand cells arrayed to form a thin spherical sheet \cite{kirkbook}. These cells 
are connected to each other by the remnants of incomplete cell division, thin membrane tubes called \emph{cytoplasmic 
bridges} \cite{kirk81_1,kirk81_2}. The ends of the cells whence emanate the flagella, however, point into the sphere 
at this stage, and so the ability to swim is only acquired once the alga turns itself inside out through an opening at the top of the 
cell sheet, called the \emph{phialopore} \cite{viamontes77,kirkreview,hallmann}.  

Of particular interest in the present context is the crucial first step of this process, the formation of a circular invagination in
so-called `type B' inversion  (Fig.~\ref{fig1}c,f) followed by the engulfing of the posterior by the anterior 
hemisphere \cite{hallmann,hohn11}.  (This scenario is distinct from `type A' inversion in which the initial steps involve four lips 
which peel back from a cross-shaped phialopore.)   The invaginations of cell sheets found in type B inversion are 
very generic deformations during morphogenetic events such as gastrulation and neurulation \cite{he,lowery,eiraku,sawyer}, 
but, in animal model organisms, they often arise from an intricate interplay of cell division, intercalation, migration and cell 
shape changes. Modelling these therefore requires cell-based models, as pioneered by Odell \emph{et al.} \cite{odell}, but 
simpler models of simpler morphogenetic processes are required to elucidate the underlying mechanics of these problems 
\cite{howard}. Inversion in \emph{Volvox} is, however, driven by active cell shape changes alone: inversion starts when 
cells close to the equator of the shell elongate and become wedge-shaped \cite{hohn11}. Simultaneously, the cytoplasmic 
bridges migrate to the wedge ends of the cells, thus splaying the cells locally and causing the cell sheet to bend \cite{hohn11}
(Fig.~\ref{fig1}d). Additional cell shape changes have been implicated in the relative contraction of one hemisphere with 
respect to the other in order to facilitate invagination \cite{hohn14}. After invagination, the bend region expands, allowing the 
posterior hemisphere to invert fully.

At a more physical level, it has been shown recently that the inversion process is simple enough to be amenable to a 
mathematical description \cite{hohn14}: the deformations of the alga are well reproduced by a simple elastic model 
in which the cell shape changes and motion of cytoplasmic bridges impart local 
variations of intrinsic curvature and stretches to an elastic shell \cite{hohn14}. The associated mechanics have remained 
unclear, however. Here, we perform an asymptotic analysis at small deformations to clarify the geometric distinction 
between deformations resulting from intrinsic bending and intrinsic stretching, respectively. A numerical study of the bifurcation 
behavior further serves to illustrate how a sequence of local deformations can achieve invagination, and how contraction 
complements bending in this picture.

\section{Elastic Model}
Following H\"ohn \emph{et al.} \cite{hohn14}, we inscribe \emph{Volvox} inversion into the very general framework of the 
axisymmetric deformations of a thin elastic spherical shell of radius $R$ and thickness $h\ll R$ under variations of its intrinsic 
curvature and stretches. The undeformed, spherical, configuration of the shell is characterized by arclength $s$ and the distance 
of the shell from its axis of revolution, $r_0(s)$ (Fig.~\ref{fig1}g). To these correspond arclength $S(s)$ and distance from the axis 
of revolution $r(s)$ in the deformed configuration (Fig.~\ref{fig1}h). The undeformed and deformed configurations are related by 
the meridional and circumferential stretches, 
\begin{equation}
f_s(s)=\frac{\mathrm{d}S}{\mathrm{d}s} \ \ \ \ \ \ {\rm and} \ \ \ \ \ \ f_\phi(s)=\frac{r(s)}{r_0(s)}~.
\label{fs_define}
\end{equation}
(These definitions do not 
require that the undeformed configuration be spherical, and apply for the deformations of any axisymmetric object.) These 
define the strains
\begin{align}
&E_s=f_s-f_s^0, &&E_\phi=f_\phi-f_\phi^0,
\end{align}
and curvature strains
\begin{align}
&K_s=f_s\kappa_s-f_s^0\kappa_s^0,&&K_\phi=f_\phi\kappa_\phi-f_\phi^0\kappa_\phi^0, 
\end{align}
where $\kappa_s$ and $\kappa_\phi$ denote the meridional and circumferential curvatures of the deformed shell. The 
intrinsic curvatures and stretches introduced by $f_s^0,f_\phi^0$ and $\kappa_s^0,\kappa_\phi^0$ extend Helfrich's work 
on membranes \cite{helfrich}. The deformed configuration of the shell minimises an energy of the Hookean 
form \cite{libai,audolypomeau,knoche11}
\begin{align}
\mathcal{E}&=\dfrac{\pi Eh}{1-\nu^2}\int_0^{\pi R}{\hspace{-3mm}r_0\Bigl(E_s^2+E_\phi^2
+2\nu E_sE_\phi\Bigr)\,\mathrm{d}s}\nonumber\\
&\hspace{5mm}+ \dfrac{\pi Eh^3}{12(1-\nu^2)}\int_0^{\pi R}{\hspace{-3mm}r_0\Bigl(K_s^2+K_\phi^2
+2\nu K_sK_\phi\Bigr)\,\mathrm{d}s}.\label{eq:enf}
\end{align}
with material parameters the elastic modulus $E$ and Poisson's ratio $\nu$. In computations, we take  
$h/R=0.15$ and $\nu=1/2$ appropriate for \emph{Volvox} inversion \cite{hohn14}.

In general, deformations of the shell arise from a complex interplay of intrinsic stretches and curvatures, and 
the global geometry of the shell. To clarify these, we begin by considering two simple kinds of deformations, in which the 
competition is between two effects only. How these effects conspire in general we shall explore in the main body of the paper.

\subsubsection{Simple Deformations: Stretching and Bending}
\begin{figure}
\includegraphics{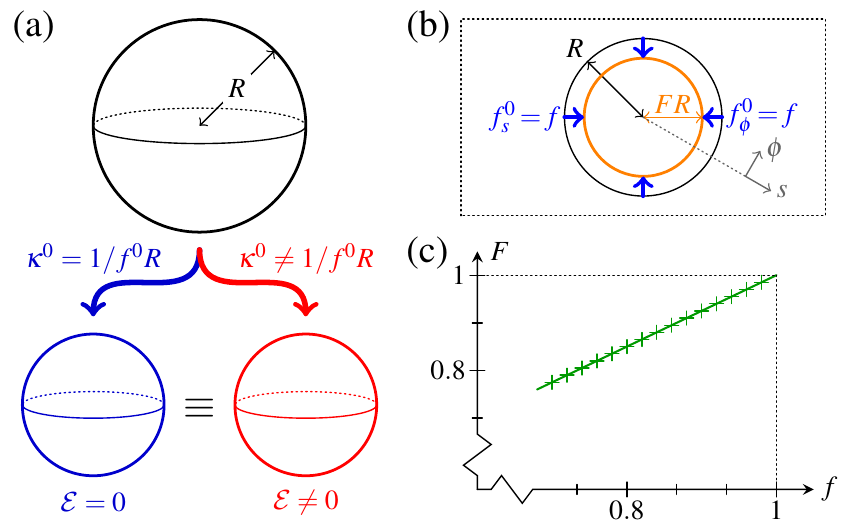}
\caption{(color online). Simple intrinsic deformations. (a)~A sphere can be shrunk to smaller spheres of equal 
radii by both compatible and incompatible intrinsic deformations. (b)~Contraction of a circular region of radius 
$R$ in a plane elastic sheet by a factor $f$. The boundary of this region is contracted to $s=FR$. (c)~Numerical 
result for $F$ (+) agrees with analytical calculation (\ref{eq:linear}) (solid line).}
\label{fig2}
\end{figure}

The simplest intrinsic deformation is one of uniform stretching or contraction, which does not affect the global, spherical 
geometry of the shell. This corresponds to $f_s^0=f_\phi^0=f$ and $\kappa_s^0=\kappa_\phi^0=1/fR$. With these intrinsic 
stretches and curvatures, the original sphere deforms to a sphere of radius $R'$. Then $f_s=f_\phi=R'/R$, and 
so the strains are $E_s=E_\phi=R'/R-f$. However, $\kappa_s=\kappa_\phi=1/R'$. Thus
$f_s\kappa_s=f_\phi\kappa_\phi=f_s^0\kappa_s^0=f_\phi^0\kappa_\phi^0=1/R$, and so $K_s=K_\phi=0$. 
The energy density is therefore proportional to $(R'/R-f)^2$ and is minimized for $R'=fR$, at which point $\mathcal{E}=0$
(Fig.~\ref{fig2}a). (Indeed, uniform contraction is a homothetic transformation: the angles between material points 
are unchanged, and so there is no bending involved. In other words, the shell is blind to its intrinsic curvature on this 
spherical solution branch.) 

The intrinsic stretches and curvatures need not be compatible in this way, however: suppose that $f_s^0=f_\phi^0=f$, 
but $\kappa_s^0=\kappa_\phi^0=1/f'R$ with $f\not=f'$. The energy still has spherical minima of radius $R'=fR$, but now 
with $\mathcal{E}\not=0$ (Fig.~\ref{fig2}a). This illustrates that, conversely, even if the equilibrium shape is spherical, the intrinsic curvatures and stretches cannot straightforwardly be inferred from the resulting shape. 

\subsubsection{Simple Deformations: Stretching and Geometry}
To illustrate how the global geometry affects these deformations, we consider contraction of a plane 
elastic sheet, with $f_s^0=f_\phi^0=f<1$ for $s<R$ (Fig.~\ref{fig2}b). This does not involve any bending 
of the sheet, and, upon non-dimensionalising lengths with $R$, the shell minimises 
\begin{align}
&\int_0^\infty{\hspace{-2mm}s\Bigl\{\bigl[r'(s)-f(s)\bigr]^2+\bigl[r(s)/s-f(s)\bigr]^2}\nonumber\\
&\hspace{15mm}+2\nu\bigl[r'(s)-f(s)\bigr]\bigl[r(s)/s-f(s)\bigr]\Bigr\}\,\mathrm{d}s, 
\end{align}
where
\begin{align}
f(s)=\left\{\begin{array}{cl}
f&\text{if }s<1\\
1&\text{if }s>1.
            \end{array}\right.
\end{align}
The resulting Euler--Lagrange equation is
\begin{align}
\dfrac{\mathrm{d}}{\mathrm{d}s}\left(s\dfrac{\mathrm{d}r}{\mathrm{d}s}\right)-\dfrac{r}{s}
=(1+\nu)(1-f)s\,\delta(s-1)\label{eq:ele}. 
\end{align}
This is a homogeneous equation, and the solution satisfying the geometric conditions $r(0)=0$ 
and $r(s)\sim s$ as $s\rightarrow\infty$ as well as continuity of $r$ at $s=1$ is
\begin{align}
r(s)= \left\{\begin{array}{cl}
Fs&\text{if }s<1\\
s+\dfrac{F-1}{s}&\text{if }s>1.
            \end{array}\right.
\end{align}
The constant $F=r(1)$ is determined by the jump condition at $s=1$, or, physically, by requiring the stress to 
be continuous across $s=1$. This finally yields
\begin{align}
F=\dfrac{1}{2}\left[(1-\nu)+(1+\nu)f\right]. 
\label{eq:linear}
\end{align}
This simplified problem serves as a test case for numerical solution of the more general Euler--Lagrange equations 
associated with (\ref{eq:enf}).  These boundary-value problems can be solved numerically 
with the solver \texttt{bvp4c} of \textsc{Matlab} (The~Mathworks, Inc.); our numerical setup of the 
governing equations otherwise mimicks that of \cite{knoche11}.  In this particular example, the linear relationship in 
(\ref{eq:linear}) is indeed confirmed numerically (Fig.~\ref{fig2}c). 
Notice that the governing equation (\ref{eq:ele}) is independent of the forcing applied away from $s=1$; the 
solution is determined by geometric boundary conditions.

\section{Results}
The most drastic cell shape changes at the start of inversion occur when cells in a narrow region close to the equator 
become wedge-shaped (Fig.~\ref{fig1}d). These are accompanied by motion of the cytoplasmic bridges to the 
thin tips of the cells to splay the cell sheet and drive its inward bending. For this reason, 
H\"ohn~\emph{et~al.} \cite{hohn14} started by considering a piecewise constant functional form for the intrinsic 
curvature, in which this curvature took negative values in a narrow region close to the equator. It was found, however, 
that with this ingredient alone the energy minimizers could not reproduce the mushroom shapes adopted by the embryos 
in the early of stages of inversion (Fig.~\ref{fig1}c,f), producing instead a shape cinched in at those points -- the
so-called `purse-string' effect.  However, analysis of thin sections had previously revealed that the cells in the 
posterior hemisphere become thinner at the start of inversion \cite{hohn11}. When the resulting contraction 
of the posterior hemisphere was incorporated into the model, it could indeed reproduce, quantitatively, the shapes of 
invaginating \emph{Volvox} embryos.  

H\"ohn \emph{et al.} have thus identified two different types of active deformations that contribute to the shapes 
of inverting \emph{Volvox} at the invagination stage: first, a localized region of active inward bending (corresponding 
to negative intrinsic curvature), and second, relative contraction of one hemisphere with respect to the other. We shall 
focus on these two types of deformation in what follows and clarify the ensuing elastic and geometrical balances. 

\begin{figure*}
\includegraphics{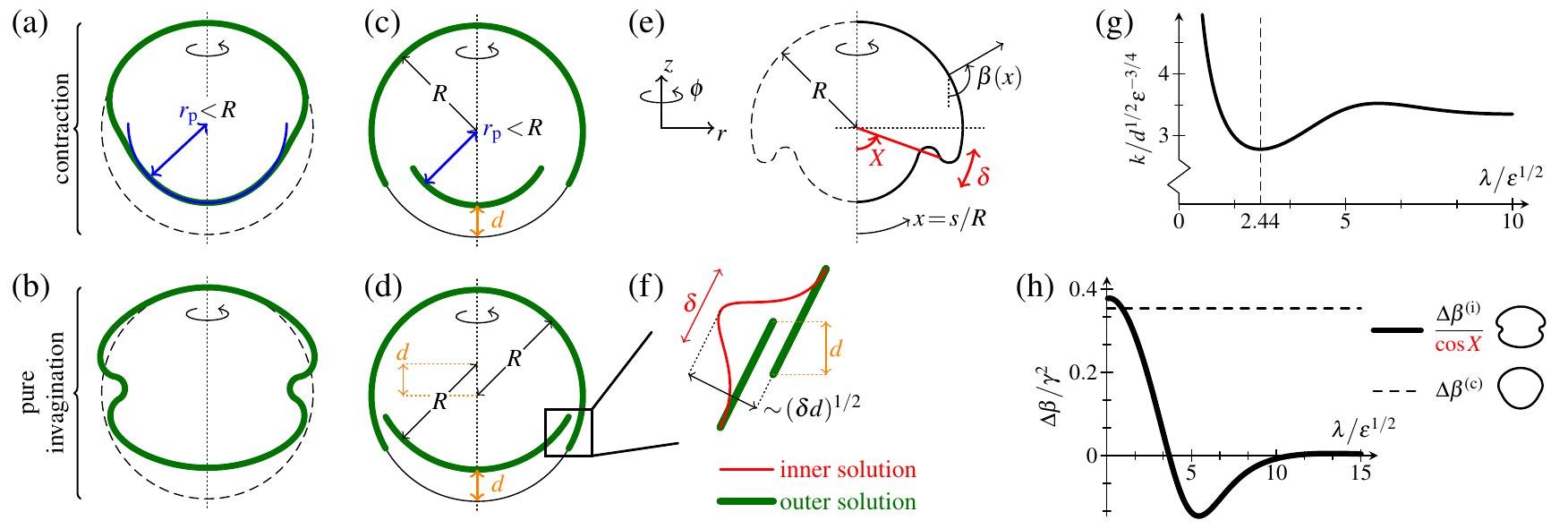}
\caption{(color online).  Asymptotic analysis of invagination and contraction. (a) Numerical shape resulting from 
contracting the posterior to a radius $r_{\mathrm{p}}<R$. (b) Numerical ``hourglass'' shape resulting from pure 
invagination. (c) Geometry of contraction with posterior radius $r_{\mathrm{p}}<R$, resulting in upward motion of the 
posterior by a distance $d$. (d) Geometry of pure invagination solution. (e) Asymptotic geometry: in the limit 
$h \ll R$, deformations are localized to an asymptotic inner layer of width $\delta$ about $x = X$, where 
$x = s/R$ is the angle that the undeformed normal makes with the vertical. In the deformed configuration, this angle 
has changed to $\beta(x)$. (f) Asymptotic invagination: upward motion of posterior by a distance $d$ requires 
inward deformations scaling as $(\delta d)^{1/2}$ in the inner layer of width $\delta$. (g) Relation between 
preferred curvature $k$ and width of invagination $\lambda$ for a given amount of upward posterior motion $d$, 
from asymptotic calculations. (h) Inward rotation of midpoint of invagination with, and without contraction, from 
asymptotic calculations.} 
\label{fig3}
\end{figure*}

\subsection{Asymptotic Analysis}
We start by seeking equilibrium configurations in the limit of a thin shell, $h\ll R$. In this limit, 
the shapes (Fig.~\ref{fig3}a,b) corresponding to contraction or (pure) invagination (by which we mean, here, 
deformations driven by a region of high intrinsic curvature only) result from the matching of spherical shells of 
different radii or disparate relative positions (Fig.~\ref{fig3}c,d). Deviations from these outer solutions 
are localized to an asymptotic inner layer of non-dimensional width $\delta$ about $x = X$, where $x = s/R$ is 
the angle that the normal to the undeformed shell makes with the vertical (Fig.~\ref{fig3}e). Here, we consider an 
incipient deformation where the normal angle $\beta(x)$ to the deformed shell deviates but slightly from its value in 
the spherical configuration, viz $\beta(x)=x+b(x)$, with $b\ll 1$.

\subsubsection{Geometric Considerations}
We begin by clarifying the geometric distinction between contraction and invagination. The radial and vertical displacements obey
\begin{subequations}
\label{eq:displ}\begin{align}
u_r'&=f_s\cos{\beta}-\cos{x}=-b\sin{X}+O\bigl(\delta b,b^2\bigr),\\
u_z'&=f_s\sin{\beta}-\sin{x}=b\cos{X}+O\bigl(\delta b,b^2\bigr),
\end{align}
\end{subequations}
where dashes denote differentiation with respect to $x$, and where we have assumed the scaling $f_s=1+O(\delta b)$ which 
we shall derive presently. Let $d$ denote the (non-dimensional) distance by which the posterior moves up. Matching 
to the outer solutions requires the net displacements $U_r$ and $U_z$, obtained by integrating (\ref{eq:displ}) 
across the inner layer, to obey
\begin{subequations}
\begin{align}
U_r^{\text{(c)}}&=d\sin{X},&U_z^{\text{(c)}}&=-d\cos{X},\label{eq:totdc}\\
U_r^{\text{(i)}}&=0,& U_z^{\text{(i)}}&=-d,\label{eq:totdnc}
\end{align}
\end{subequations}
where the superscripts (c) and (i) refer, respectively, to the solutions corresponding to contraction and (pure) invagination. 
In the case of contraction, (\ref{eq:displ}) and (\ref{eq:totdc}) give the scaling $b^{\text{(c)}}\sim d/\delta$. If there is 
no contraction, however, (\ref{eq:displ}) and (\ref{eq:totdnc}) imply that the leading-order solution does not yield 
any upward motion of the posterior, which is associated with a higher-order solution only. This suggests that the 
appropriate scaling is $b^{\text{(i)}}\sim(d/\delta)^{1/2}$, which we shall verify presently. 

Our assumption $b\ll 1$ thus translates to $d\ll\delta$. Hence, in the invagination case, upward motion of the posterior 
requires comparatively large inward displacements of order $(\delta d)^{1/2}\gg d$ (Fig. \ref{fig3}f). This asymptotic 
difference of the deformations corresponding to contraction and invagination arises purely from geometric effects; it 
is the origin of the `purse-string' shapes found by H\"ohn \emph{et al.} in the absence of 
contraction \cite{hohn14}.

\subsubsection{Elasto-Geometric Considerations}
Here, we discuss the detailed solution for pure invagination. Upon non-dimensionalising distances with $R$ and stresses 
with $Eh$, the Euler--Lagrange equations of~(\ref{eq:enf}), derived in the appendix, can be cast into the form
\begin{subequations}
\label{eq:gov}\begin{align}
&f_s\Sigmai\sin{x}\tan{\beta}-\varepsilon^2\cos{\beta}\Bigl(1-\nu\beta'-\sin{\beta}\cosec{x}\Bigr)\nonumber\\
&\hspace{6mm}-\varepsilon^2\dfrac{\mathrm{d}}
{\mathrm{d}x}\Bigl(\beta'\sin{x}+\nu\bigl(\sin{\beta}-\sin{x}\bigr)\Bigr)=k^0(x),\\
&\dfrac{\mathrm{d}}{\mathrm{d}x}\Bigl(\Sigmai\sec{\beta}\sin{x}\Bigr)-A-\nu\Sigmai=0,
\end{align}
\end{subequations}
with the small parameter
\begin{align}
\varepsilon^2=\dfrac{1}{12(1-\nu^2)}\dfrac{h^2}{R^2}\ll 1. 
\end{align}
In these equations, $\Sigmai$ is the non-dimensional meridional stress, and $A=e_\phi$ is the dimensionless hoop strain. 
The contribution from the intrinsic curvature is
\begin{equation}
k^0(x) = \varepsilon^2\left\{\nu\kappa_s^0(x)\cos{\beta}
-\dfrac{\mathrm{d}}{\mathrm{d}x}\Bigl(\kappa_s^0\sin{x}\Bigr)\right\}. 
\end{equation}
The equations are closed by the geometric relation
\begin{align}
\label{eq:geo}\dfrac{\mathrm{d}}{\mathrm{d}x}\Bigl(A\sin{x}\Bigr)=f_s\cos{\beta}-\cos{x}. 
\end{align}
Introducing $\gamma=(d/\delta)^{1/2}$, scaling gives the leading balances 
$\Sigmai\sim\varepsilon^2\gamma/\delta^2$, $\Sigmai/\delta\sim A$, and $A/\delta\sim\gamma$ in 
(\ref{eq:gov},\ref{eq:geo}). Hence $\delta\sim\varepsilon^{1/2}$, and we define an inner coordinate $\xi$ via 
$x=X+\delta\xi$. We also introduce the expansions
\begin{subequations}
\begin{align}
\beta &= X+\gamma\bigl(b_0+\gamma b_1+\gamma^2b_2+\cdots\bigr),\\
A &= \delta\gamma\bigl(a_0+\gamma a_1+\cdots\bigr),\\
\Sigmai &= \delta^2\gamma\cot{X}\bigl(\sigma_0+\gamma\sigma_1+\cdots\bigr).
\end{align}
\end{subequations}
This further proves the scaling $f_s=1+O(\delta\gamma)$ that we have assumed previously. 

The pure invagination configuration is forced by intrinsic curvature that differs from the curvature of the undeformed 
sphere in a region of width $\lambda$ about $x=X$, where $\kappa_s^0=-k$. 
Writing $\Lambdai=\lambda\big/\varepsilon^{1/2}$, we thus have, at leading order,
\begin{equation}
\kappa_s^0(\xi)=-\dfrac{d^{1/2}}{\varepsilon^{3/4}}K\Bigl(\mathrm{H}\bigl(\xi+\tfrac{1}{2}\Lambdai\bigr) 
- \mathrm{H}\bigl(\xi-\tfrac{1}{2}\Lambdai\bigr)\Bigr),
\end{equation}
where $k=d^{1/2}\varepsilon^{-3/4}K$, and where H denotes the Heaviside function. Thus
\begin{align}
&k^0(\xi) =\varepsilon^{3/4}d^{1/2}K^0(\xi)\sin{X}, 
\end{align}
where
$K^0(\xi)=K\left[\delta\left(\xi+\tfrac{1}{2}\Lambdai\right) - \delta\left(\xi-\tfrac{1}{2}\Lambdai\right)\right]$.
We note that $\gamma^3\gg\gamma\delta$ provided that $d\gg\varepsilon$ (which we shall assume to be the case); thus 
we may set $x=X$ to the order at which we are working. Expanding (\ref{eq:gov},\ref{eq:geo}), we then find
\begin{align}
&\sigma_0-b_0''=K^0(\xi),&&\sigma_0'-a_0=0,&&a_0'=-b_0, 
\end{align}
at lowest order, where dashes now denote differentiation with respect to $\xi$. At next order,
\begin{subequations}
\begin{align}
&\sigma_1+\sigma_0b_0\sec{X}\cosec{X}-b_1''=0,\\
&\sigma_1'+\dfrac{\mathrm{d}}{\mathrm{d}\xi}\bigl(b_0\sigma_0)\sin{X}\sec{X}-a_1=0,
\end{align}
\end{subequations}
with 
 $a_1'=-b_1-\tfrac{1}{2}b_0^2\cot{X}$.
We are left to determine the matching conditions by expanding (\ref{eq:displ}) to find
\begin{subequations}
\begin{align}
u_r' &= -\delta\gamma b_0\sin{X}-\delta\gamma^2\bigl(b_1\sin{X}+\tfrac{1}{2}b_0^2\cos{X}\bigr)\nonumber\\
&\hspace{6mm}-\delta\gamma^3\bigl(b_2\sin{X}+b_0b_1\cos{X}-\tfrac{1}{6}b_0^3\sin{X}\bigr),\\
u_z' &= \delta\gamma b_0\cos{X}+\delta\gamma^2\bigl(b_1\cos{X}-\tfrac{1}{2}b_0^2\sin{X}\bigr)\nonumber\\
&\hspace{6mm}+\delta\gamma^3\bigl(b_2\cos{X}-b_0b_1\sin{X}-\tfrac{1}{6}b_0^3\cos{X}\bigr),
\end{align}
\end{subequations}
up to corrections of order $O\bigl(\delta\gamma^4,\delta^2\gamma\bigr)$. Applying (\ref{eq:totdnc}), at lowest order, we find
\begin{equation}
\int_{-\infty}^\infty{b_0\di\xi}=0. 
\label{eq:match1}
\end{equation}
At next order, (\ref{eq:totdnc}) is a system of two linear equations for two integrals, with solution
\begin{align}
&\int_{-\infty}^\infty{b_0^2\di\xi}=2\sin{X},&&\int_{-\infty}^\infty{b_1\di\xi}=-\cos{X}. 
\label{eq:match2}
\end{align}
We note in particular that the resulting condition on the leading-order solution has only arisen in the second-order 
expansion of the matching conditions. Similarly, at order $O\bigl(\delta\gamma^3\bigr)$, we find
\begin{equation}
\int_{-\infty}^\infty{b_0b_1\di\xi}=0. \label{eq:sym}
\end{equation}
The leading-order problem is thus
\begin{equation}
b_0''''+b_0=K\Bigl(\delta''\bigl(\xi+\tfrac{1}{2}\Lambdai\bigr) - \delta''\bigl(\xi-\tfrac{1}{2}\Lambdai\bigr)\Bigr),\label{eq:lo}
\end{equation}
with matching conditions (\ref{eq:match1}) and the first of (\ref{eq:match2}).
Symmetry ensures that the first of (\ref{eq:match1}) is satisfied. After a considerable amount of algebra, 
the first of (\ref{eq:match2}) reduces to a relation between $K$ and $\Lambdai$,
\begin{equation}
K^2=\dfrac{8\sqrt{2}\sin{X}}{1+\e^{-\Lambdai/\sqrt{2}}\Bigl(\bigl(\sqrt{2}\Lambdai-1\bigr)
\sin{\frac{\Lambdai}{\sqrt{2}}}-\cos{\frac{\Lambdai}{\sqrt{2}}}\Bigr)}. 
\end{equation}
This function exhibits a global minimum at $\Lambdai\approx 2.44$ (Fig.~\ref{fig3}g). This is a first indication that 
narrow invaginations are more efficient than those resulting from wider regions of high intrinsic curvature, 
a statement that we shall make more precise later. 

Symmetry also implies that there is no inward rotation of the midpoint of the invagination at this order. Rather, inward 
folding is a second-order effect, for which we need to consider the second-order problem,
\begin{equation}
b_1''''+b_1=\left\{\dfrac{\mathrm{d}^2}{\mathrm{d}\xi^2}\Bigl(b_0\sigma_0\Bigr)-\tfrac{1}{2}b_0^2\right\}\cot{X},
\label{eq:so}
\end{equation}
with matching conditions (\ref{eq:sym}) and the second of (\ref{eq:match2}).
The rotation of the midpoint of the invagination is thus
\begin{equation}
\Deltai\beta^{\text{(i)}}=\Bigl(B^{\text{(i)}}(\Lambdai)\cos{X}\Bigr)\dfrac{d}{\varepsilon^{1/2}}, \label{eq:midr}
\end{equation}
where $B^{\text{(i)}}(\Lambdai)$ is determined by the solution of (\ref{eq:so}). The detailed solution reveals that
\begin{widetext}
\begin{equation}
B^{\text{(i)}}(\Lambdai)=\frac{2\sqrt{2}\e^{-\frac{\Lambdai}{2 \sqrt{2}}} \left[4 \e^{\frac{\Lambdai}{2 \sqrt{2}}}
\sin{\frac{\Lambdai }{\sqrt{2}}}-\e^{\frac{\Lambdai}{\sqrt{2}}} \sin{\frac{\Lambdai }{2 \sqrt{2}}}-3 \sin{\frac{3 \Lambdai }
{2 \sqrt{2}}}+\Bigl(\e^{\frac{\Lambdai}{\sqrt{2}}}-1\Bigr) \cos{\frac{\Lambdai }{2 \sqrt{2}}}\right]}
{5 \left[\e^{\frac{\Lambdai}{\sqrt{2}}}+\bigl(\sqrt{2} \Lambdai -1\bigr) \sin{\frac{\Lambdai }{\sqrt{2}}}-\cos{\frac{\Lambdai}
{\sqrt{2}}}\right]},
\end{equation}
\end{widetext}
but the geometric factor in (\ref{eq:midr}) is the main point: this factor resulting from the global geometry of the shell 
hampers the inward rotation of the midpoint of the invagination. (This is as expected: by symmetry, invagination at the equator, 
where $\cos{X}=0$, yields no rotation.) 

An analogous, though considerably more straightforward, calculation can be carried out for contraction: 
non-dimensionally, upward posterior motion by $d$ requires $f_s^0=f_\phi^0=1-d$ for $x<X$, and leads to 
\begin{align}
\Deltai\beta^{\text{(c)}}=\dfrac{1}{2\sqrt{2}}\dfrac{d}{\varepsilon^{1/2}}.
\end{align}
At this order, the above solutions for pure invagination and contraction can be superposed; in particular, the 
solutions at order $O\bigl(\gamma^2\bigr)$ have the same symmetry, and so (\ref{eq:sym}) is satisfied. For contraction, there 
is thus no geometric obstacle to inward folding (Fig.~\ref{fig3}h). Contraction is thus not only a means 
of creating the disparity in the radii of the anterior and posterior hemispheres required to fit the partly inverted latter into 
the former, but also drives the inward folding of the invagination, by breaking its symmetry. In \emph{Volvox} inversion, 
this symmetry breaking is at the origin of the formation of the second passive bend region highlighted by H\"ohn 
\emph{et al.} \cite{hohn14} to stress the non-local character of these deformations.

\begin{figure}
\includegraphics{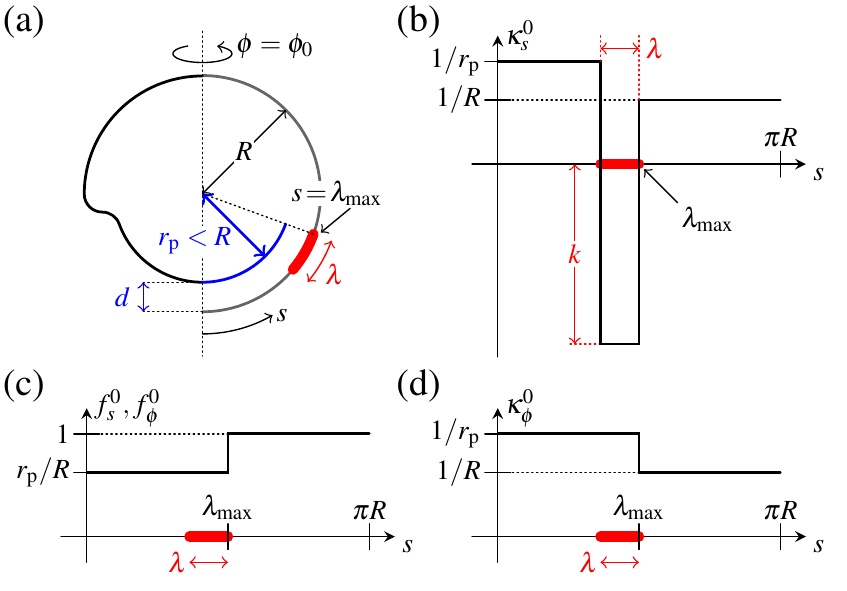}
\caption{(color online). Setup for numerical calculations, following \cite{hohn14}. (a)~Geometrical setup: the intrinsic 
curvature $\kappa_s^0$ of a spherical shell of undeformed radius $R$ differs from the undeformed curvature in the 
range $\lambda_{\max} > s > \lambda_{\max}-\lambda$, where $s$ is arclength. Posterior contraction is taken into 
account by a reduced posterior radius $r_\mathrm{p}<R$. These intrinsic curvature and contraction result in deformations 
that move up the posterior pole by a distance~$d$. (b)~Corresponding functional form of $\kappa_s^0$; in the bend 
region, $\kappa_s^0=-k<0$. (c)~Form of the intrinsic stretches $f_s^0,f_\phi^0$ for posterior contraction. 
(d)~Functional form of $\kappa_\phi^0$ for posterior contraction.} 
\label{fig4}
\end{figure}

\subsection{Bifurcation Behaviour}
The asymptotic analysis has shown that the coupling of elasticity and geometry constrains small invagination-like 
deformations both locally and globally, but that contraction can help overcome these global constraints. These ideas 
carry over to larger deformations of the shell, which must however be studied numerically. For this purpose, we extend 
the setup of \cite{hohn14}, motivated by direct observation of thin sections of fixed embryos: the intrinsic 
curvature $\kappa_s^0$ differs from that of undeformed sphere in the 
range $\lambda_{\max}>s>\lambda_{\max}-\lambda$ of arclength along the shell (Fig.~\ref{fig4}a). In this region of 
length $\lambda$, $\kappa_s^0=-k$, where $k>0$ (Fig.~\ref{fig4}b). This imposed intrinsic curvature results in 
upward motion of the posterior pole by a distance $d$.

\subsubsection{Stability Statements}

\begin{figure}
\includegraphics{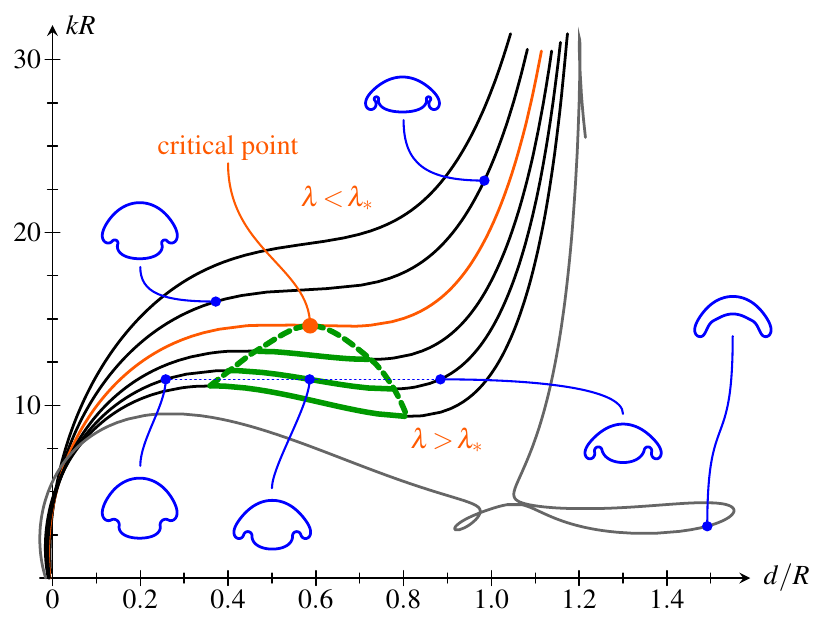}
\caption{(color online).  Bifurcation behaviour of invagination solutions. Solution space for $\lambda_{\max} = 1.1R$
and $r_{\mathrm{p}}=R$: each line shows the relation between $k$ and $d$ at some constant $\lambda$. A critical 
branch (at $\lambda=\lambda_*$) separates different types of branches. Branches with $\lambda>\lambda_\ast$ feature 
two extrema; the resulting spinodal curve (thick dashed line) defines a critical point. Insets illustrate representative 
solution shapes. See text for further explanation.} 
\label{fig5}
\end{figure}

Our first observation is that, at fixed $\lambda_{\max}$, more than one solution may arise for the same input parameters 
$(k,\lambda)$. Further understanding is gained by considering, at fixed $\lambda_{\max}$ and for different 
values of $\lambda$, the relation between $k$ and $d$. The typical behaviour of these branches is plotted in 
Fig.~\ref{fig5}. (The shapes eventually self-intersect; accordingly, these branches end but we expect 
them to be joined up smoothly to configurations with opposite sides of the shell in contact. The study of such 
contact configurations typically requires some simplifying assumptions to be made \cite{knoche11}, but we do 
not pursue this further, here.)

\begin{figure*}
\includegraphics{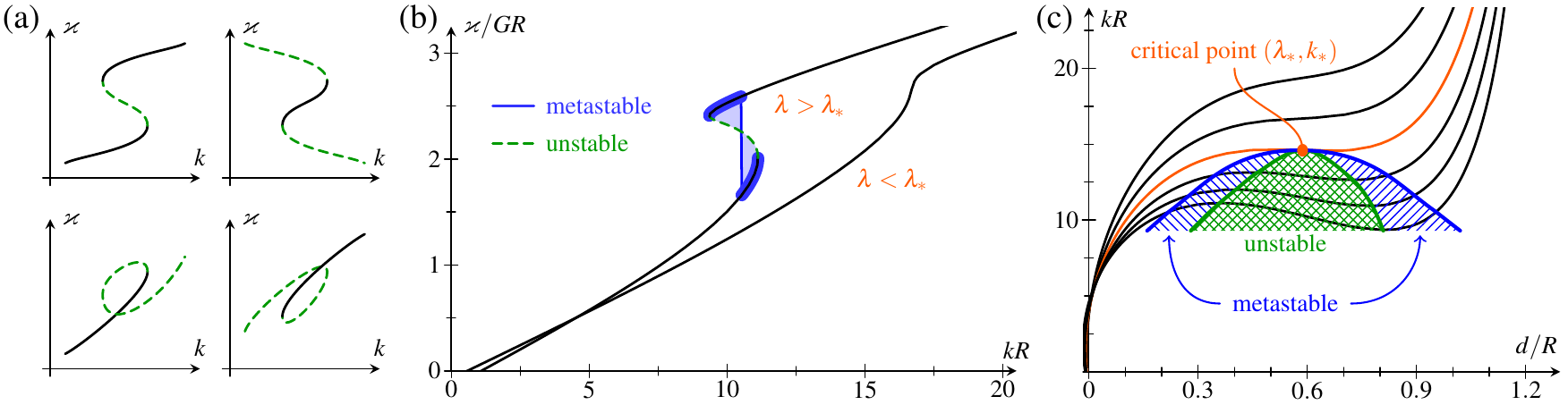}
\caption{(color online). Stability of invagination solutions. (a)~Possible topologies of a double fold in the distinguished 
$(k,\varkappa)$ bifurcation diagram. Dashed branches are those that the results of \cite{maddocks87} imply to be 
unstable. (b)~For $\lambda>\lambda_\ast$, S-shaped folds arise in the $(k,\varkappa)$ diagram. From general 
theory \cite{maddocks87}, the middle part of the branch is unstable, while the outer parts are stable. An additional 
region of metastability is identified by the Maxwell construction. (c)~Resulting picture: a region of unstable and 
metastable solutions expands underneath the critical point.} 
\label{fig6}
\end{figure*}

At the distinguished value $\lambda=\lambda_\ast$, a critical branch arises (Fig.~\ref{fig5}). It separates two types of 
branches: first, those  with $\lambda<\lambda_\ast$, on which $d$ varies mononotonically with $k$, and second, those 
with $\lambda>\lambda_\ast$, where the relation between $d$ and $k$ is more complicated. At large values of $\lambda$, 
these branches may have a rather involved topology involving loops. At values of $\lambda$ just above $\lambda_\ast$, 
however, there is a range of values of $k$ for which there exist three configurations (Fig.~\ref{fig5}). We note that the two 
outer configurations have $\partial d/\partial k>0$, while the middle one has $\partial d/\partial k<0$. The latter 
behaviour prefigures instability, which we shall discuss in more detail below. There are thus two points on these branches 
where $k$, viewed locally as a function of $d$, reaches an extremum. The curve joining up these extrema for different 
values of $\lambda$ we shall term the `spinodal curve'. This curve, in turn, has a maximum at a point on the 
critical branch, which we shall call the `critical point' and which is characterised by~$\lambda_\ast$ and the critical 
curvature, $k_\ast$.

The stability of the configurations in Fig.~\ref{fig5} can be assessed by means of general results of
bifurcation theory \cite{maddocks87}, used recently to discuss the stability of the buckled equilibrium shapes of 
a pressurised elastic spherical shell \cite{knoche11,knoche14}. If we let $\varkappa=-\partial\mathcal{E}/\partial k$ denote 
the conjugate variable to $k$, the key result of \cite{maddocks87} is that stability, at fixed $\lambda$, of extremizers of 
the energy $\mathcal{E}$ can be assessed from the folds in the $(k,\varkappa)$ bifurcation diagram. In particular, stability 
can only change at folds in the bifurcation diagram. Expanding the bending part of the energy functional (\ref{eq:enf}) 
for $f_s^0=f_\phi^0=1$ and $\kappa_\phi^0=0$, we find
\begin{equation}
\varkappa = -G\int_{\lambda_{\max}-\lambda}^{\lambda_{\max}}{r_0\Bigl(f_s\kappa_s
+\nu f_\phi\kappa_\phi+k\Bigr)\,\mathrm{d}s}
\end{equation}
with $G=\pi Eh^3\big/6\bigl(1-\nu^2\bigr)$. (The last term in the integrand is independent of the solution, and 
may therefore be ignored in what follows.)

The two folds that arise in the $(k,d)$ diagram for $\lambda>\lambda_\ast$ (Fig.~\ref{fig5}) are compatible 
\emph{a priori} with four fold topologies in the $(k,\varkappa)$ diagram (Fig.~\ref{fig6}a). However, since a single 
solution exists for small $k$ (at fixed $\lambda$), the lowest branch must be stable. Further, since the branches do 
not self-intersect in the $(k,d)$ diagram, they cannot self-intersect in the $(k,\varkappa)$ diagram either. The results 
of \cite{maddocks87} imply that only the first topology in Fig.~\ref{fig6}a is compatible with this, and so the fold is 
S-shaped and traversed upwards in the $(k,\varkappa)$ diagram. It follows in particular that the middle branch, 
with $\partial d/\partial k<0$ is unstable, and that right branch is stable. (Numerically, one confirms that the branches 
are indeed S-shaped.) Thus the stability of the branches in this simple bifurcation diagram could also be inferred from 
the $(k,d)$ diagram (though, in general problems, as discussed in \cite{maddocks87}, different bifurcation diagrams 
may suggest contradictory stability results). However, the Maxwell construction of equal areas \cite{landaulifshitzsp} 
can be applied to the $(k,\varkappa)$ diagram (Fig.~\ref{fig6}b) to identify metastable solutions beyond the unstable 
branch. These stability considerations may appear rather technical, but they are in fact very natural: under reflection, 
the $(d,k)$ diagram maps to the diagram of isotherms of a classical van-der-Waals gas, for which the middle branch 
is well known to be unstable \cite{landaulifshitzsp}. Under this analogy, $\mathcal{E}$ corresponds to the Gibbs free 
energy of the gas.

This analysis cannot immediately be extended to the more exotic topologies that arise for $\lambda$ close to 
$\lambda_{\max}$ (Fig.~\ref{fig5}). We note however that part of these branches must be unstable, too: as above, 
a single solution exists for small $d$, and so the corresponding branch must be stable. The first fold must be traversed 
upwards, and the first branch with $\partial d/\partial k<0$ is thus unstable, as above.

An analogous analysis can be carried out for deformations that vary $\lambda$ while keeping $k$ fixed: for $k>k_\ast$, 
the $(d,\lambda)$ diagram is monotonic, but this ceases to be the case for $k<k_\ast$. As above, the stability 
can be inferred from the $(\lambda,d)$ diagram, and the middle branch with $\partial d/\partial \lambda<0$ is unstable, too.

The picture that emerges from this discussion is the following: solutions in a region of parameter space underneath 
the critical point bounded by the spinodal curve are unstable; a band of solutions on either side of this region and 
below the critical point are metastable (Fig.~\ref{fig6}c), both to perturbations varying $k$ and to perturbations 
varying $\lambda$. If invagination, driven by a localized region of active bending, is to be stable, it must move 
around the critical point: if it were to enter the unstable region, the shell would flip back and forth between the 
`shallow' and `deep' invagination states on either side of the unstable region and suffer large strains in the 
process. (This makes this kind of instability different from the classical buckling instability of a rod or a 
`popper' toy \cite{pandey}: the latter is directed in that, once the instability threshold is crossed, the system will snap to the new 
preferred configuration and remain there.) The need for a sequence of stable deformations to move around the 
critical point rationalises the timecourse of invagination in \emph{Volvox}: initially, a narrow band of cells 
undergoes cell shape changes, thereby acquiring a high intrinsic curvature. This region of cells then widens, 
moving around the critical point, whereupon the preferred curvature relaxes and posterior inversion can complete.

\subsubsection{Contraction and Criticality}
\begin{figure}
\includegraphics{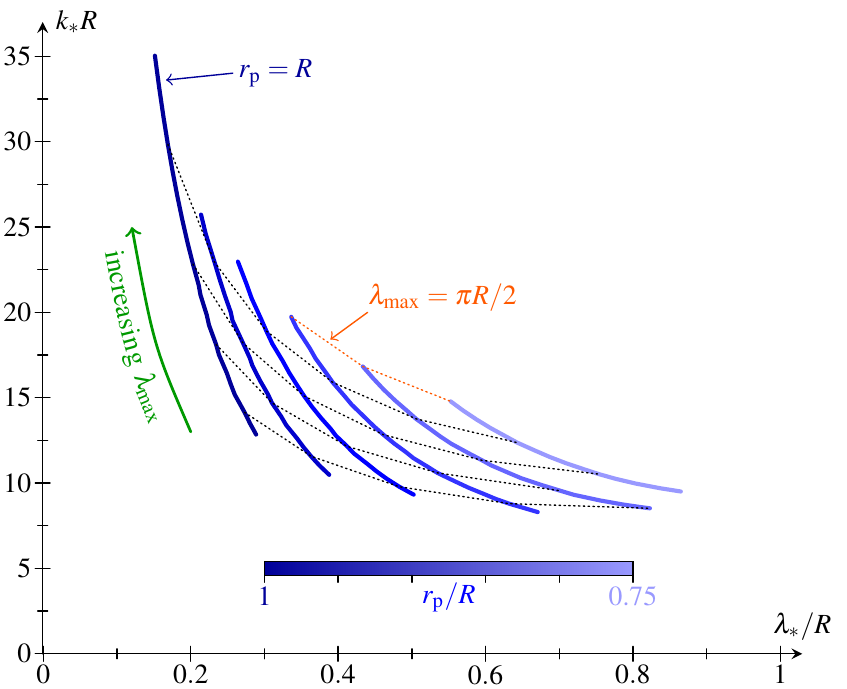}
\caption{(color online). Contraction and the critical point. Trajectories of critical point in parameter space as $\lambda_{\max}$ is varied, for different values of $r_{\mathrm{p}}$. Thin dotted lines are curves of constant $\lambda_{\max}$. At constant $\lambda_{\max}$, increased contraction leads to decreased $k_\ast$ and increased $\lambda_\ast$.} 
\label{fig7}
\end{figure}

For different values of $\lambda_{\max}$, the critical point traces out a trajectory in parameter space, characterised 
by $k_\ast$ and $\lambda_\ast$ (Fig.~\ref{fig7}). As $\lambda_{\max}$ increases, $k_\ast$ increases, while 
$\lambda_\ast$ decreases. Thus the closer to the equator, the more difficult invagination is, not only because there is 
less room to fit the posterior into the anterior, but also because a stable invagination requires narrower and 
narrower invaginations of higher and higher intrinsic curvature.  

We are left to explore how contraction affects the position of the critical point, and hence the invagination. We introduce 
a reduced posterior radius $r_{\mathrm{p}}<R$ as in~\cite{hohn14} (Fig.~\ref{fig4}a), and modify the intrinsic curvatures 
and stretches accordingly (Fig.~\ref{fig4}b,c,d). Numerically, we observe that, at constant $\lambda_{\max}$, 
increasing contraction (that is, reducing $r_{\mathrm{p}}$) decreases the critical curvature $k_\ast$, and 
increases $\lambda_\ast$ (Fig.~\ref{fig7}). Hence contraction aids invagination not only geometrially, but also mechanically: 
first, it allows invagination close to the equator (which would otherwise be prevented by different parts of the 
shell touching), and second, it makes stable invagination easier, by reducing $k_\ast$. Thus, again, contraction 
appears as a mechanical means to overcome global geometric constraints.

\subsubsection{Asymptotic Analogy}
In the asymptotic analysis in the previous section, we restricted ourselves to small deviations of the normal angle from 
the spherical configuration so that the problem remained analytically tractable. While the leading scaling balances remain 
the same for large rotations, the resulting non-linear ``deep-shell equations'' cannot be rescaled so that the dependance 
on $X$ drops out \cite{audolypomeau}. Some further insight can, however, be gained in the shallow-shell limit 
$X\ll 1$: in terms of the inner coordinate $\xi$, we write
\begin{align}
\beta(\xi)=X\,B(\xi) \ \ \ \ {\rm and} \ \ \ \  \Sigmai(\xi)=\varepsilon\,S(\xi).
\end{align}
In the absence of forcing by intrinsic curvature or stretches, the leading-order balance is
\begin{align}
2S''=1-B^2 \ \ \ \ {\rm and} \ \ \ \ B''=SB,\label{eq:shallow}
\end{align}
where dashes denote, as before, differentiation with respect to $\xi$.

This balance arises also in the study of a spherical shell pushed by a plane \cite{audolypomeau}: at large 
indentations, the shell dimples and the plane remains in contact with it only in a circular transition region 
joining up the undeformed shell to the isometric dimple. With the matching conditions $B\rightarrow\pm 1$ 
as $\xi\rightarrow\pm\infty$, (\ref{eq:shallow}) describe the leading-order shape of this transition region 
\cite{audolypomeau}. Remarkably, this deformation is independent of the contact force, which only arises at the 
next order in the expansion \cite{audolypomeau}. 

\begin{figure}
\includegraphics{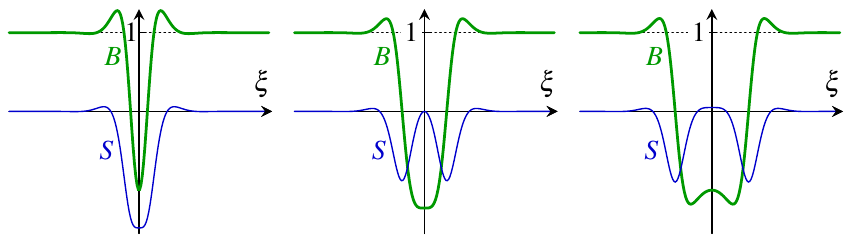}
\caption{(color online). Examples of ``preferred'' deformation modes, which are solutions of (\ref{eq:shallow}). 
Of the modes shown, the middle one has the lowest elastic energy.} 
\label{fig8}
\end{figure}

The appropriate boundary conditions for the invagination case are $B\rightarrow 1$ as $\xi\rightarrow\pm\infty$, and 
non-constant solutions of (\ref{eq:shallow}) can indeed be found numerically (some solutions are shown in Fig.~\ref{fig8}). 
In these modes, the deformations are, in a sense, large compared to intrinsic curvature imposed, making them 
geometrically `preferred'. Their existence lies at the heart of the bifurcation behaviour discussed above.

\section{Conclusion}
In this paper, we have explored perhaps the simplest intrinsic deformations of a spherical shell: elastic and geometric 
effects conspire to constrain deformations resulting from a localized region of intrinsic bending. Contraction, a somewhat 
more global deformation, alleviates these constraints and thereby facilitates the stable transition from one configuration of 
the shell to another. This rich mechanical behaviour makes a mathematically interesting problem in its own right, yet 
this analysis has implications for \emph{Volvox} inversion and wider material design problems.

Experimental studies of \emph{Volvox} inversion \cite{hohn11,hohn14} had revealed the existence of posterior 
contraction, and indeed, the simple elastic model that underlies this paper can only reproduce \emph{in vivo} shapes 
once posterior contraction is included \cite{hohn14}. Of course, contraction is an obvious means of creating a disparity 
in the anterior and posterior radii required ultimately to fit one hemisphere into the other, but the present analysis reveals 
that, beyond this geometric effect, there is another, more mechanical side to the coin: if contraction is present, lower 
intrinsic curvatures, i.e. less drastic cell shape changes, are required to stably invert the posterior hemisphere. This ascribes 
a previously unrecognized additional role to these secondary cell shape changes (i.e. those occurring away from the main 
bend region): just as the shape of the deformed shell arises from a glocal competition between elastic and geometric 
effects, a combination of local and more global intrinsic properties allow inversion to proceed stably. Thus, as we have 
pointed out previously, this mechanical analysis rationalises the timecouse of the observed cell shape changes, thereby 
lending further support to the observation of H\"ohn \emph{et al.} \cite{hohn14}, that it is a spatio-temporally well 
regulated sequences of cell shape changes that drives inversion. Thus, the remarkable process of \emph{Volvox} inversion 
is mechanically more subtle than it may initially appear to be.

Intrinsic deformations that allow transitions of an elastic object from one configuration to another are of inherent interest 
in the material design context, and divide into two classes: first, snapping transitions for fast transitions between states, 
studied in \cite{bende}, and second, stable sequences of intrinsic deformations. The glocal behaviour of the latter 
is illustrated by the present analysis: in particular, additional transformations such as contraction can 
increase the number of stable parameter paths between configurations of the elastic object. In this material design 
context, non-axisymmetric deformations such as polygonal folds or wrinkles \cite{vella11} could also become important, and 
may warrant a more detailed analysis.

\section*{Acknowledgements}
We thank Stephanie H\"ohn, Aurelia R. Honerkamp-Smith and Philipp Khuc Trong for extensive discussions. This work 
was supported in part by an
EPSRC studentship (PAH), an EPSRC Established Career Fellowship (REG), and a Wellcome Trust Senior Investigator Award
(REG).

\bigskip
\section*{Appendix: Governing Equations}
\setcounter{equation}{0}
\renewcommand{\theequation}{A\arabic{equation}}
In this appendix, we sketch the derivation of the Euler--Lagrange equations of the energy functional (\ref{eq:enf}), following \cite{knoche11}. The variation takes the form
\begin{align}
\dfrac{\delta\mathcal{E}}{2\pi}&=\int_0^{\pi R}{\hspace{-3mm}r_0\Bigl(N_s\,\delta E_s+N_\phi\,\delta E_\phi\Bigr)\di s}\nonumber\\
&\qquad\qquad+\int_0^{\pi R}{\hspace{-3mm}r_0\Bigl(M_s\,\delta K_s+M_\phi\,\delta K_\phi\Bigr)\di s},\label{eq:evar}
\end{align}
where we have introduced the stresses and moments
\begin{subequations}
\begin{align}
N_s&=C\bigl(E_s+\nu E_\phi\bigr), &N_\phi&=C\bigl(\nu E_s+E_\phi\bigr),\\
M_s&=D\bigl(K_s+\nu K_\phi\bigr),&M_\phi&=D\bigl(\nu K_s+K_\phi\bigr),
\end{align}
\end{subequations}
with $C=Eh\big/\bigl(1-\nu^2\bigr)$ and $D=Ch^2/12$. (These stresses and moments are expressed here relative to the undeformed configuration.) 

The deformed shape of the shell is characterised by the radial and vertical coordinates $r(s)$ and $z(s)$, as well as the angle $\beta(s)$ that the normal to the deformed shell makes with the vertical direction. These geometric quantities obey the equations \cite{knoche11}
\begin{align}
&\d{r}{s}=f_s\cos{\beta},&&\d{z}{s}=f_s\sin{\beta},&&\d{\beta}{s}=f_s\kappa_s\label{eq:geom}.
\end{align}
We note that one of these is redundant. The variations $\delta E_s,\delta E_\phi,\delta K_s,\delta K_\phi$ are purely geometrical, and one shows that \cite{knoche11}
\begin{subequations}
\begin{align}
\delta E_s&=\sec{\beta}\,\delta r'+f_s\tan{\beta}\,\delta\beta,&\delta E_\phi&=\dfrac{\delta r}{r_0},\\
\delta K_s&=\delta\beta',&\delta K_\phi&=\dfrac{\cos{\beta}}{r_0}\delta\beta.
\end{align}
\end{subequations}
The variation (\ref{eq:evar}) thus becomes
\begin{widetext}
\begin{align}
\dfrac{\delta\mathcal{E}}{4\pi}&= \Bigl\llbracket r_0N_s\sec{\beta}\,\delta r+r_0M_s\,\delta\beta\Bigr\rrbracket-\int_0^{\pi R}{\biggl\{\d{}{s}\Bigl(r_0N_s\sec{\beta}\Bigr)-N_\phi\biggr\}\delta r\di s}\nonumber\\
&\hspace{60mm}+\int_0^{\pi R}{\biggl\{r_0f_sN_s\tan{\beta}+M_\phi\cos{\beta}-\d{}{s}\Bigl(r_0M_s\Bigr)\biggr\}\delta\beta\di s},\label{eq:evar2}
\end{align}
\end{widetext}
upon integration by parts, whence
\begin{subequations}
\begin{align}
&r_0f_sN_s\tan{\beta}+M_\phi\cos{\beta}-\d{}{s}\Bigl(r_0M_s\Bigr)=0,\\
&\d{}{s}\Bigl(r_0N_s\sec{\beta}\Bigr)-N_\phi=0.
\end{align}
\end{subequations}
These equations, together with two of the geometric relations (\ref{eq:geom}), describe the shape of the deformed shell. 
For numerical purposes, it is convenient to remove the singularity at $\beta=\pi/2$ by introducing the transverse shear 
tension \cite{libai,knoche11}, $Q=-N_s\tan{\beta}$, expressed here relative to the undeformed configuration. Force balance 
  arguments \cite{libai,knoche11} show that $Q$ obeys 
\begin{align}
\d{}{s}\Bigl(r_0Q\Bigr)+r_0f_s\kappa_sN_s+r_0f_\phi\kappa_\phi N_\phi =0.
\end{align}
The solution $Q=-N_s\tan{\beta}$ is selected by the boundary condition $Q(0)=0$. At the poles of the shell, the equations 
have singular terms in them, but these singularities are either removable or the appropriate boundary values are set by 
symmetry arguments \cite{knoche11}.  This allows appropriate boundary conditions and values to be derived.

\end{document}